\documentclass[11pt,a4paper]{article}
\usepackage{jcappub}

\title{Bailing Out the Milky Way: Variation in the Properties of Massive Dwarfs Among Galaxy-sized Systems}

\author{Chris W. Purcell}
\author{and Andrew R. Zentner}

\affiliation{Department of Physics and Astronomy \& 
Pittsburgh Particle physics, Astrophysics and Cosmology Center (PITT PACC), 
University of Pittsburgh 15260 USA}

\emailAdd{cpurcell@pitt.edu}
\emailAdd{zentner@pitt.edu}

\abstract{
Recent kinematical constraints on the internal densities of the Milky Way's 
dwarf satellites have revealed a discrepancy with the subhalo populations of simulated 
Galaxy-scale halos in the standard cold dark matter model of hierarchical structure formation.  
In particular, the Via Lactea II and Aquarius simulations both have large subhalos with 
internal densities that are larger than the constraints inferred for any Milky Way dwarf satellites. 
This has been dubbed the ``too big to fail'' problem, with reference to the improbability 
of large and invisible companions existing in the Galactic environment.
In this paper, we argue that both the Milky Way observations and simulated subhalos are 
consistent with the predictions of the standard model for structure formation.  Specifically, we 
show that there is significant variation in the properties of subhalos among distinct host halos of 
fixed mass and suggest that this can reasonably account for the deficit of dense satellites in the Milky Way.  
We exploit well-tested analytic techniques to predict the properties in a large sample of distinct host halos 
with a variety of masses spanning the range expected of the Galactic halo.  Such techniques render the 
problem of estimating the variance in subhalo properties computationally feasible.  The analytic model 
produces subhalo populations consistent with both Via Lactea II and Aquarius, and our results suggest 
that natural variation in subhalo properties suffices to explain the discrepancy between Milky Way 
satellite kinematics and these numerical simulations.  At least $\sim10\%$ of Milky Way-sized halos 
host subhalo populations for which there is no ``too big to fail'' problem, even when the host halo mass 
is as large as $M_{\mathrm{host}} = 10^{12.2}\,h^{-1}$~M$_{\odot}$.  Follow-up studies consisting of 
high-resolution simulations of a large number of Milky Way-sized hosts are necessary to 
confirm our predictions.  In the absence of such efforts, the ``too big to fail'' problem does not appear 
to be a significant challenge to the standard model of cold dark matter halos undergoing hierarchical 
formation.
}


\arxivnumber{1208.4602}

\begin{document}
\maketitle
\flushbottom

\section{Introduction} 

In the standard paradigm describing structure growth in our universe, galaxies such as 
the Milky Way form within halos of dark matter \cite{white_rees78,blumenthal_etal84}.   
This formation is hierarchical in the sense that small, self-bound clumps of dark matter 
collapse first into roughly virialized objects, subsequently merging at the nodes of sheets 
and filaments of dark matter to form ever larger halos.  Most of the merging halos are disrupted 
within a few dynamical times and are no longer recognizable as distinct objects within the larger 
halo; however, some survive to the present day within the virialized regions of their host halos.  
The properties of these dark subhalos and their luminous components, such as the satellite galaxies of the 
Milky Way, are probes of the faintest substructure surrounding the Milky Way and 
may serve as sensitive testbeds of cosmological structure formation  
\cite{liddle_kamionkowski00,zb03,tollerud_etal08,bullock_etal10,font_etal11,rashkov_etal12}.

High-resolution simulations of $\Lambda$CDM cosmological evolution have been able to 
model the properties of subhalos within host halos similar to the Milky Way.  However, the 
computational expense involved in such an undertaking has necessarily limited the number 
of experiments in which dwarf-sized satellites can be easily resolved at the ultra-faint scale 
($L_{\mathrm{dwarf}} \gtrsim 10^5$~L$_{\odot}$) to less than a dozen (taking as representative 
the set of Aquarius, GHALO, and Via Lactea II simulations in refs.~\cite{springel_etal08}~\cite{stadel_etal09}, 
and \cite{diemand_etal07}, respectively).  Consequently, the diversity of subhalo characteristics and their 
natural variation among different host halos at fixed mass, as well as the details of those properties' 
dependence on halo mass, have only been subject to a relatively small number of studies.

Recent theoretical results in the context of Milky Way dwarf galaxy constraints suggest 
that the densest, most massive subhalos found in the Via Lactea II and Aquarius simulations have properties 
that are not commensurable with those of observed satellites in the Local Group \cite{BK_etal11}.
This builds upon the early indications and possible cosmological implications of such a mismatch explored in 
ref.~\cite{zb03}, and raises the question of whether the densest, most massive dwarf galaxies predicted by 
simulations are present in the real Milky Way environment. If these few largest halos are present, 
then why are they not luminous like their slightly less massive counterparts?  In a properly cosmological 
context, it would seem that these putative satellites are ``too big to fail", meaning that they are sufficiently more 
massive than the other Milky Way satellites that it is difficult to construct models in which these satellites 
remain invisible \cite{BK_etal11}. In such a scenario, these subhalos are either dense and somehow dark 
relics \cite{BK_etal11}, or have had their baryonic content scoured via early feedback effects 
\cite{diCintio_etal11,BK_etal12,zolotov_etal12}.  An alternative to these astrophysical solutions would conclude 
that these few most massive subhalos are not to be found and that this discrepancy is truly reflective of a critical 
failure in the $\Lambda$CDM structure formation paradigm (as proposed, for example, in the self-interacting CDM 
model of \cite{vogelsberger_etal12}).  

In this work, we explore another possible solution to the ``too big to fail" problem.  We use analytic models to 
argue that the natural, statistical variation in subhalo densities from one host halo to another is significant and 
can affect the theoretical interpretation of satellite galaxy data.  In particular, we suggest that the properties of 
the few densest, most massive Milky Way satellites do not pose a challenge to galaxy or structure formation 
theories.  Rather, we argue that the observed satellite properties are consistent with being randomly drawn from 
among the statistical variety of subhalo populations that can be realized in Milky Way-sized halos, with a probability 
that is not negligible.  

Halo-to-halo variation in subhalo properties is a natural consequence of the hierarchical structure-formation model 
as subhalos merge to form the contemporary host halo through an infinitude of distinct mass accretion histories.  
Any particular Galactic-scale halo is attended by a subhalo population with properties that are functions of this assembly 
history and likely a number of other variables.  At present, we have some ability to constrain viable formation routes of 
the Milky Way.  For example, our constraints on several properties of the thin and dynamically-cold Galactic disk imply 
that the host halo is unlikely to have suffered a minor merger (with mass ratio $\lesssim 10:1$) passing through the disk within 
the past 5-8 Gyr \cite{hammer_etal07,purcell_etal09}, before the ongoing Sagittarius impact and the imminent absorption 
of the Magellanic Clouds end that quiescent history at the present day \cite{law_etal10,purcell_etal11,besla_etal07,busha_etal11}.  
However, existing constraints are broad and cannot be used to limit effectively the properties of the subhalos that should surround 
the Milky Way.  Moreover, we do not have a comprehensive understanding of which variables are most important in determining the 
properties of the Milky Way subhalo population.  

At the same time, we have only a poor handle on the Milky Way's total mass because stellar rotation curves only probe the central 
regions of the Milky Way halo and other probes yield a wide range of values \cite{xue_etal08,li_white08,przybilla_etal10,busha_etal11}.  
Interestingly and unfortunately, this mass range is fraught with cosmologically-significant transitions that bear directly on 
galaxy formation \cite{codis_etal12,pichon_etal11,guo_etal10}, and numerical investigations have suggested that host halo 
mass is the largest determinant of subhalo properties in this particular context, with gas cooling and other baryonic effects taking a 
secondary role \cite{geen_etal12}.  In addition, a recent statistical exploration of subhalo abundances in a simulated $\Lambda$CDM 
universe has illustrated the sensitivity of satellite populations to host halo mass, finding that Galaxy-sized halos with mass 
$M_{\mathrm{host}} \lesssim 10^{12} $~M$_{\odot}$ are much more likely to have only three large companions (analogous to the Large 
and Small Magellanic Clouds and the disrupting Sagittarius dwarf) at the present-day \cite{wang_etal12}.  

We complement this and other numerical efforts with our ability to probe host-mass parameter space, and most importantly the intrinsic 
variation among subhalo properties at fixed host-mass, in much greater detail due to the computational advantages of our analytic approach.  
In the remainder of the paper, we itemize our methods and results.  We describe the details of our formalism in \S~\ref{sec:methods}, 
including a restatement in \S~\ref{sec:toobig} of the ``too big to fail'' discrepancy in terms of a new variable describing central subhalo 
density.  We give our basic results, culminating in probability distributions for this density-proxy variable, in \S~\ref{sec:results}, 
reserving \S~\ref{sec:discuss} for discussion and interpretation in the context of future numerical experiments as well as achievable 
(and useful) observational constraints.

\section{Methods}
\label{sec:methods}

Our investigations complement existing results in their focus on the quantification of the natural scatter in subhalo properties 
from one host halo to the next.  In order to make such an estimate, it is necessary to derive the properties of subhalos in a very 
large number of host halos, a challenging task with direct numerical simulations because of the computational costs of such an 
effort.  We overcome this difficulty by using an analytic technique to predict the properties of subhalo populations that approximates 
the results of numerical simulations at greatly reduced computational cost.  The technique has been described in detail in 
\cite{zentner_etal05}, to which we refer the reader for further details, including a description of the algorithm as well as a 
demonstration that the model predictions for subhalo abundances, internal structures, and spatial distributions within host halos 
agree well with high-resolution numerical experiments in the regimes where the two techniques are commensurable.  
Several related techniques have been developed that produce results that are broadly similar but differ in some detail 
\cite[e.g.][]{zb03,zentner07,gan_etal10,yang_etal11,taylor_babul04,taylor_babul05a,taylor_babul05b,giocoli_etal10}.
Similar methods have been applied to estimate the variance in putative dark matter annihilation signals 
yielded by galactic substructures \cite{koushiappas_etal10}.

The analytic calculations produce subhalo populations within individual, statistically distinct, 
host halos of a given mass.  We refer to each host halo as a {\em realization}, because the 
subhalo properties within any given host are determined by a particular realization of the 
density field in the host environment.  For each realization, the technique produces a catalog 
of subhalos, each being described by a variety of properties including: time at which the subhalo 
first merged into the host; subhalo mass at merger; final subhalo mass after evolving within the 
host to the present day; internal subhalo structure; and orbital position and velocity.  These properties 
will enable us to make the estimates of interest to the present manuscript, since we can quantify the 
natural variation in subhalo properties among hosts by producing a very large number of distinct 
realizations at fixed host halo mass.  

We work within a cosmological model described by the best-fitting parameters from 
the seven-year analysis of the Wilkinson Microwave Anisotropy Probe team \cite{komatsu_etal11}.  
We note that a halo's virial mass is calculated inside the region for which 
the mean density is greater than the mean matter density of the universe by a factor $\Delta$, that 
evolves in redshift from $\Delta(z=0) \sim 337$ to $\Delta(z) \sim 178$ for $z > 1$, as in the fitting algorithm 
of ref.~\cite{bryan_norman98}.Throughout this work, we define subhalos as those self-bound objects inside 
this virial radius $R_{\mathrm{vir}}$ at the present day.  
We examine three host halo masses that span an interesting range compared to both 
current very high-resolution numerical simulations, such as Via Lactea II and 
Aquarius, as well as current bounds on the halo mass of the Milky Way: 
$M_{\mathrm{host}}=10^{11.8}\, h^{-1}$~M$_{\odot}$, $M_{\mathrm{host}}=10^{12.0}\, h^{-1}$~M$_{\odot}$, 
and $M_{\mathrm{host}}=10^{12.2}\,h^{-1}$~M$_{\odot}$.  This mass range spans the 
various estimates for the present-day virial mass $M_{\mathrm{host}}$ 
of the Milky Way, yielded by a number of methods including but not limited to 
the dynamical timing of Local Group galaxies and satellites of M31 
\cite[$M_{\mathrm{host}} \simeq 10^{12.15}~h^{-1} $~M$_{\odot}$]{li_white08,vdM_etal12}, 
as well as extrapolations drawn from the rotation curve of stars in the 
Galactic stellar halo \cite[$10^{11.85 \pm 0.1}~h^{-1} $~M$_{\odot}$]{xue_etal08} and statistical 
comparisons with cosmological simulation results regarding the brightest Galactic satellites 
\cite[$10^{11.92 \pm 0.2}~h^{-1} $~M$_{\odot}$]{busha_etal11}.  
The Via Lactea II halo has a mass of $M_{\mathrm{host}} \simeq 10^{12.2}\,h^{-1}$~M$_{\odot}$ 
\cite[][hereafter VL2]{vl2_nature}.  In the Aquarius suite, six simulated halos span 
the mass range from $M_{\mathrm{host}} \simeq 10^{12}\, h^{-1}$~M$_{\odot}$ to 
$M_{\mathrm{host}} \simeq 10^{12.2}\, h^{-1}$~M$_{\odot}$ \cite{springel_etal08}, and we note 
that Aquarius subhalos are identified within a radius defined by the virial overdensity $\Delta = 250$ 
(and therefore a larger radius than all three models considered here, in which 
$R_{\mathrm{vir}} \sim 171, 271, 429$~kpc in order of ascending mass).  Of these Aquarius subhalos, 
the analysis of ref.~\cite{BK_etal11} adopts an outer limit of $R < 300$ kpc for comparison to observed 
Galactic dwarfs.

At each of the three host halo masses that we model, we generate 10,000 realizations of halo 
substructure.  These subhalo populations enable us to make an approximate assessment 
of the variation in satellite properties among different host halos at fixed host halo mass.  
Additionally, this large number of realizations enables us to sub-divide our samples while 
still making statistically-significant statements.  For example, we explore cuts on subhalo accretion 
redshift and other properties in order to approximate some of the specific conditions that may pertain 
to the present state of the Milky Way disk, and the large size of our total realization set results in 
subsamples that are each statistically valuable in terms of Poisson error.  

Before closing this section, we remind the reader that our technique is 
an approximation to direct simulations.  The technique generally compares 
well to simulation results; however, as only a relatively small number of 
simulations exist in which substructure in Milky Way-sized halos is 
very well resolved, it is not possible to rule out moderate systematic 
errors associated with this technique.   Furthermore, this technique 
cannot be validated outside the range of scales that are well resolved 
by numerical simulations, and like the simulations to which we 
compare, it treats only the evolution of dark matter 
halos and neglects baryonic evolution.  While the influence of 
baryons on subhalo structure on these scales is thought to be moderate 
at most, this remains a possible source of systematic error.  
These are important caveats, and they demand that 
the results of this study must be verified by future numerical 
simulation campaigns.  Nevertheless, approximate techniques are presently 
the only way to estimate halo-to-halo variation in subhalo population 
properties at reasonable computational cost.

\begin{figure}[!]
\includegraphics[width=6.0in]{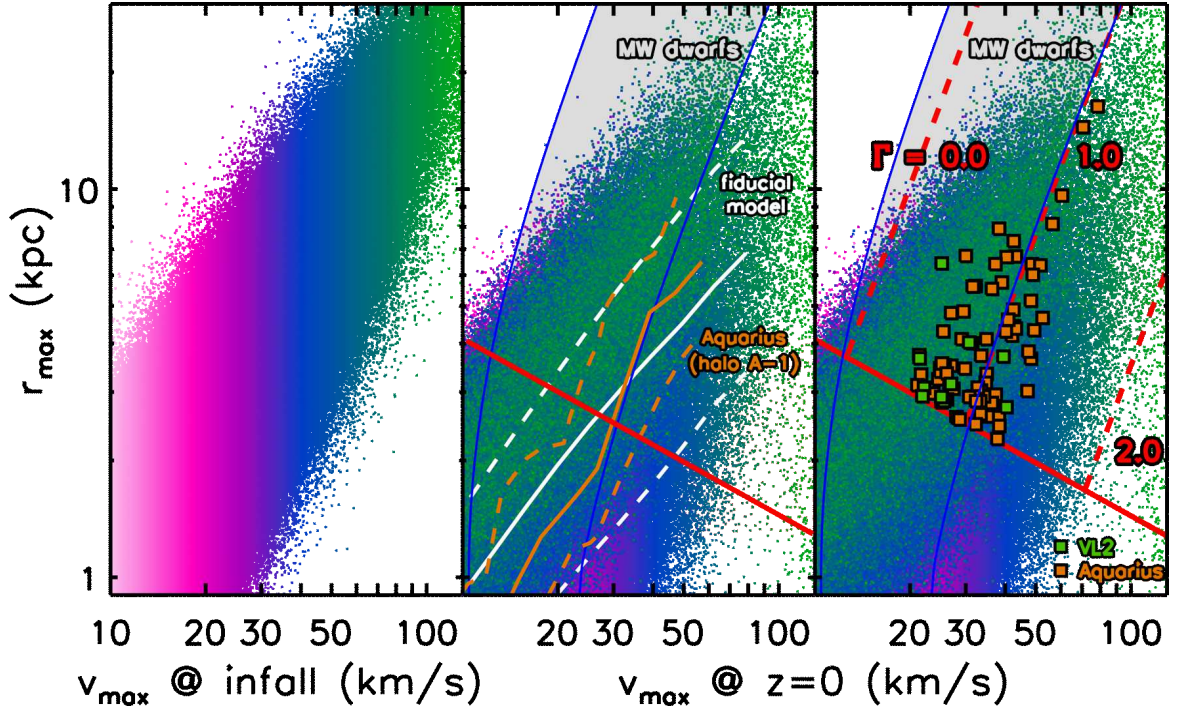}
\caption{The distribution of all satellites in 10,000 realizations of a Galactic-scale host halo with total mass $M_{\mathrm{host}} = 10^{12.0}~h^{-1} $~M$_{\odot}$, 
in the $v_{\mathrm{max}} - r_{\mathrm{max}}$ parameter space.  As shown in the {\em left} panel, all subhalo points are colored by their value of $v_{\mathrm{max}}$ 
at infall through the virial radius of the host.  In the {\em center} panel, we show in {\em white solid} and {\em dashed} lines the mean and $68\%$-range of our fiducial 
$v_{\mathrm{max}} - r_{\mathrm{max}}$ distribution at redshift $z=0$ in comparison to the subhalo mean and $68\%$-range for Aquarius halo A-1 (as published in 
\cite{springel_etal08}; shown in {\em orange}).  Finally, in the {\em right} panel, we develop a parametrization fixed by the $2\sigma$-constraint on the kinematic structures 
of the nine classical dwarfs around the Milky Way (denoted by the {\em gray shaded} region, as in \cite{BK_etal11}; see also \cite{Wolf_etal10}).  The subhalo distributions 
in modern simulations of Milky Way-like host halos, including the VL2 simulation and all Aquarius subhalos in this regime, are shown in the {\em right} panel for satellites 
in the size regime where the observational envelope resembles a power-law, and we introduce the parameter $\Gamma$ describing the relevant direction of increasing 
density in the space of $\log(v_{\mathrm{max}})$-$\log(r_{\mathrm{max}})$.  
}
\label{fig:rvmax}
\end{figure}

 \section{Reformulating the ``Too Big To Fail'' Problem}
\label{sec:toobig}

The ``too big to fail'' problem, as stated by \cite{BK_etal11}, is based on the claim that the most massive 
subhalos in simulations are too dense to be consistent with existing constraints on the classical Galactic 
dwarf galaxies.  Here, the internal densities of satellites are characterized in the space of $v_{\mathrm{max}}$ and 
$r_{\mathrm{max}}$, which are proxies for mass and size, respectively.  As usual, we define $v_{\mathrm{max}}$ 
as the peak circular velocity in the satellite's rotation curve, which occurs at a radius $r_{\mathrm{max}}$ along 
that curve.  These two parameters suffice to specify the subhalo density profile, provided it follows the typical 
Navarro-Frenk-White form \cite{nfw96}.  In the standard picture of galaxy formation, more massive subhalos are expected to host brighter 
dwarf galaxies.  The typical conclusion drawn in the literature thus far: if these massive and overly-dense satellites 
represent a generic prediction of $\Lambda$CDM, then either something is badly wrong with our 
cosmological and astrophysical models, or there really are large and {\em invisible} dwarf galaxies around the Milky Way.  
The latter statement is at odds with the basic predictions of galaxy formation theory, since such 
massive systems would seem to be ``too big to fail'' in the baryonic sense.

Stellar kinematical data can yield a precise constraint on the subhalo masses of the Milky Way dwarf 
galaxies at particular radii \cite{Wolf_etal10}.  These data constrain the structural parameters 
$v_{\mathrm{max}}$ and $r_{\mathrm{max}}$, and these constraints can be compared to the properties of 
subhalos in catalogs obtained from high-resolution numerical simulations of Galaxy-scale host halos 
(VL2 and Aquarius, as in ref. \cite{BK_etal11}), or our analytic host halo realizations.  
However, the data best constrain a degenerate combination of $v_{\mathrm{max}}$ and $r_{\mathrm{max}}$ 
\cite[a point emphasized in][]{zb03}.  The comparison of data with theoretical predictions is, 
therefore, a comparison along one particular dimension in the $\log(v_{\mathrm{max}})$-$\log(r_{\mathrm{max}})$ 
plane.  A more detailed comparison may be possible, but the multidimensional parameter constraints on 
$v_{\mathrm{max}}$ and $r_{\mathrm{max}}$ for each dwarf are not available from ref.~\cite{BK_etal11}.  

To make an approximate comparison of data with theory along the relevant dimension in the two-dimensional 
parameter space, we introduce a new parameter $\Gamma$, which represents the linear combination of 
$\log(v_{\mathrm{max}})$ and $\log(r_{\mathrm{max}})$ perpendicular to the line of $\log(r_{\mathrm{max}})$-$\log(v_{\mathrm{max}})$ 
degeneracy.  Using the contours plotted in Figs.~1-2 of ref.~\cite{BK_etal11}, we define the 
parameter $\Gamma \equiv 1+\mathrm{log}(0.0014v_{\mathrm{max}}^{2.2}/r_{\mathrm{max}})$, 
which increases in a direction that is approximately orthogonal to the envelope of constraint on Milky Way 
dwarfs, simply expressing the formulation derived in ref.~\cite{Wolf_etal10}.  We normalize the scale of $\Gamma$
such that $\Gamma = 1.0$ along the upper $2\sigma$-bound of the allowable Milky Way dwarf region 
in the $v_{\mathrm{max}} - r_{\mathrm{max}}$ parameter space, and lower (higher) values of $\Gamma$ describe 
satellites that are less (more) dense and therefore less (more) discrepant with the few most massive Milky Way dwarfs.  
Roughly speaking, subhalos with $\Gamma > 1$ are significantly denser than the observed Milky Way dwarfs.  
These are the satellites that are ``too big to fail'' according to the present literature \cite{BK_etal11}.  

In Figure~\ref{fig:rvmax}, we illustrate the basic degeneracy of $v_{\mathrm{max}}$ and $r_{\mathrm{max}}$ and 
introduce our definition of $\Gamma$, by showing the $v_{\mathrm{max}} - r_{\mathrm{max}}$ space that contains 
the massive subhalos in all 10,000 realizations of the host halo of mass $M_{\mathrm{host}} = 10^{12.0}\,h^{-1}$~M$_{\odot}$ 
(our most similar model to the Aquarius A-1 halo for which the subhalo mean and scatter in $v_{\mathrm{max}} - r_{\mathrm{max}}$ have 
been published in \cite{springel_etal08}), at their infall into the virial radius of the host (in the leftmost panel) and the same population 
at the present day (in the center and right panels, where we develop the observationally-motivated constraint on Milky Way 
populations and the associated parameter $\Gamma$ defined above).  We will compare predictions to data by evaluating 
distributions of the values of the parameter $\Gamma$ for subhalos with the Milky Way dwarf satellite data from \cite{BK_etal11}.  
As shown in Fig.~\ref{fig:rvmax}, our subhalo populations generally agree with the Aquarius A-1 result in terms of mean and scatter, 
although in our region of interest involving relatively high-mass subhalos at the present-day, we sample many accretion events that 
are more rare by orders of magnitude than any surviving subhalos in Aquarius.  These massive satellites are the rough counterparts to 
the large Galactic dwarfs such as the Sagittarius dwarf and the Magellanic Clouds, and we show that in general, these subhalos will 
evolve the most dramatically in $v_{\mathrm{max}} - r_{\mathrm{max}}$ space, as their tidal mass-loss proceeds more efficiently than 
lower-mass satellites, which as a distribution remain relatively stationary.

The relevant direction indicated by $\Gamma$ is not a fundamental statement, because it involves observational data and will change 
as they evolve (or $\Gamma$ may cease to be useful altogether with significantly more powerful constraints).  Moreover, a proper 
statistical accounting of the ``too big to fail" problem would include the fact that each observed dwarf galaxy has a unique degeneracy 
direction in the $v_{\mathrm{max}} - r_{\mathrm{max}}$ plane by individually weighting each observational constraint and related error.  
We place such an effort beyond the scope of our present work, until future observations more clearly define these degeneracy directions 
as well as each individual dwarf's $\Gamma$-value; even with such data, only an extensive search of parameter space could identify 
individual realizations that are potentially analogous to the Milky Way in more detail.  The basic goal of our analysis is to assess the relevance of the 
``too big to fail" problem as it is now typically constructed in the literature, {\em i.e.} the abundance of numerically-predicted subhalos 
outside the aggregated two-sigma envelope defined by observational constraints on Galactic dwarf satellites.  We therefore adopt a 
parametrization that addresses this specific boundary, as a rough indicator of statistical consistency.

\begin{figure}
\includegraphics[width=6.0in]{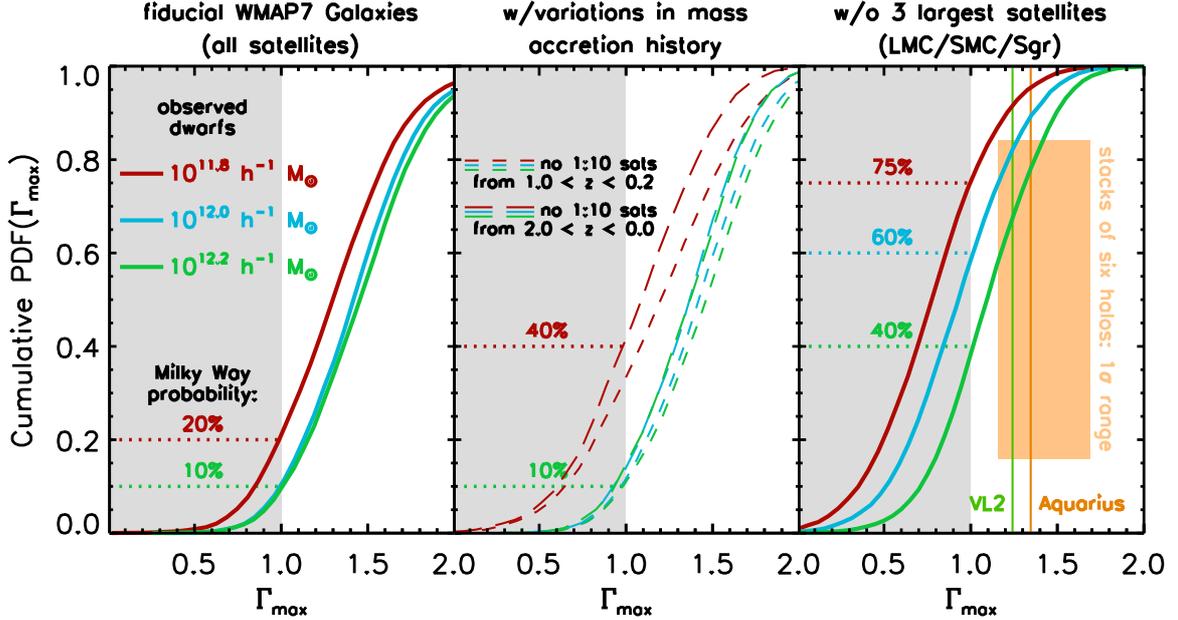}
\caption{Cumulative probability-density functions, across all 10,000 realizations in each modeled host halo mass, describing the 
distribution of the largest subhalo-value per realization of the parameter $\Gamma_{\mathrm{max}}$, as described in \S~\ref{sec:results}.  
$\Gamma_{\mathrm{max}} \le 1$ is our gross criterion for determining whether or not a particular subhalo 
population is consistent with the observed properties of the Milky Way dwarfs, hence the {\em gray shaded} region in all panels.  
In the {\em left} panel, we include all subhalos and evaluate the probability of drawing a dwarf galaxy population consistent with the 
observed Milky Way satellites.  The {\em center} panel explores the effect of cutting the fiducial samples, 
requiring a relatively quiescent accretion history for inclusion (see relevant text in \S~\ref{sec:results}.  Following the example of \cite{BK_etal11}, 
in the {\em right} panel we remove from our fiducial samples the three densest subhalos, most similar to the presently-ongoing accretions of the 
Magellanic Clouds and the Sagittarius dwarf.  The {\em orange shaded} region in this panel represents the mean $\pm 1\sigma$
range of $\Gamma_{\mathrm{max}} = (1.37\pm0.21,1.51\pm0.18)$~for halo masses $M_{\mathrm{host}} = (10^{12.0}, 10^{12.2})~h^{-1} $~M$_{\odot}$, 
where realizations are stacked in multiples of six for comparison to the Aquarius suite of halos also residing in this mass range. }
\label{fig:gamma}
\end{figure}

 \section{Results}
\label{sec:results}

Having defined the relevant dimension in the $v_{\mathrm{max}} - r_{\mathrm{max}}$ space in terms of 
the parameter $\Gamma$, which increases in the direction perpendicular to the upper-$2\sigma$ boundary of the 
Milky Way subhalo densities, we consider the region shown in Figure~\ref{fig:rvmax} above which this 
bound closely resembles a power law.  This corresponds roughly to satellites with 
$v_{\mathrm{max}} \gtrsim 30$~km/s at infall into the Galactic halo (or equivalently $r_{\mathrm{max}} \gtrsim 1-3$~kpc); 
in general, subhalos evolve toward lower values of $v_{\mathrm{max}}$ as they lose mass due 
to tidal stripping after infall (and consequently also toward lower values of $r_{\mathrm{max}}$), 
and they do so approximately along the direction of degeneracy in kinematic solutions to the Milky Way satellites.  
As Fig.~\ref{fig:rvmax} demonstrates, the region of parameter space in which $0 \lesssim \Gamma \lesssim 1$ is amply 
populated by subhalos from our analytic models that would be consistent with the Galactic constraints.  By comparison, 
massive subhalos in the VL2 and Aquarius simulations (appearing as shown in the analysis of \cite{BK_etal11}) are 
broadly consistent with the results of our analytic models, while also moderately populating the observationally-allowed 
region of the $v_{\mathrm{max}} - r_{\mathrm{max}}$ plane.  

Any individual realization of a subhalo population is a particular sub-sample of this broad distribution of satellites.  In order to make a simple 
comparison between the data and theoretical predictions, we characterize the subhalo population within a given host halo realization by a 
single parameter and explore the distribution of this parameter among host halo realizations.  We select the maximum value of $\Gamma$ in 
any given realization as our summary statistic, $\Gamma_{\mathrm{max}}$.  If $\Gamma_{\mathrm{max}} \le 1$, then {\em all subhalos} in that 
realization would lie within the observationally-allowed region of the $v_{\mathrm{max}}$-$r_{\mathrm{max}}$ parameter space.  In the absence 
of posterior distributions in $v_{\mathrm{max}}$ and $r_{\mathrm{max}}$ for each of the observed dwarf galaxies, it is difficult to make a more 
detailed comparison.  We adopt $\Gamma_{\mathrm{max}} \le 1$ as our gross criterion intended to describe satellite populations consistent 
with the Milky Way dwarf satellites as constrained by ref.~\cite{Wolf_etal10} and adopted by ref.~\cite{BK_etal11}.

In Figure~\ref{fig:gamma}, we show the cumulative probability density for a random Milky Way realization to 
have a specific value of $\Gamma_{\mathrm{max}}$, allowing us to determine what fraction of host halos have 
a subhalo population where the maximally-dense satellite has $\Gamma_{\mathrm{max}} \le 1.0$.  Including 
all satellites and all realizations, the probability of meeting this constraint is $\sim 10-20\%$, increasing as modeled 
host mass decreases.  This already suggests that the ``too big to fail'' discrepancy is not a severe problem, even if 
stated in its most constraining form.  Rather, the lack of observed dwarf satellites with ``high densities'' (high values 
of $\Gamma_{\mathrm{max}}$) may well be accommodated within the standard model of cosmological structure formation, 
given the variation in subhalo populations from one host halo to the next.  

Beyond this comparison, it is interesting to explore the influence of particular cuts on 
halo formation on the distribution of $\Gamma_{\mathrm{max}}$ and how any such cuts may 
influence our interpretation of the evolution of the Milky Way in light of the 
``too big to fail'' conjecture.  We are motivated by observational clues that the Galactic assembly 
history has been relatively quiescent over the past several Gyr \cite{hammer_etal07}, and by 
theoretical work that has shown the present-day Milky Way to be incompatible with any 
significant mass accretion events during that time \cite{purcell_etal09}.  
If we thus impose any reasonable kind of relatively quiescent mass accretion history, as shown in the two such cuts 
displayed in the center panel of Fig.~\ref{fig:gamma}, the probability of randomly drawing a Galactic satellite population 
increases to $\sim 10-40\%$ for our range of host masses.  Specifically, we identify the accretion-event scale of interest 
by the mass ratio $M_{\mathrm{host}}:M_{\mathrm{sat}} \lesssim 10:1$, in accordance with the rough constraints derived from the 
thin and dynamically-cold state of the Milky Way disk.  The most 
stringent cut we impose requires that no such event has occurred since $z=2$, while the less severe cut eliminates $10:1$ events 
since $z=1$ while also allowing for recent minor mergers after $z=0.2$.  In all halo-mass cases, increasing the quiescence of the 
assembly history only fractionally increases the probability (according to the $\Gamma_{\mathrm{max}}$ analysis) that a Galaxy-analog 
realization will be consistent with the Milky Way satellite constraints.

\begin{figure}
\includegraphics[width=6.0in]{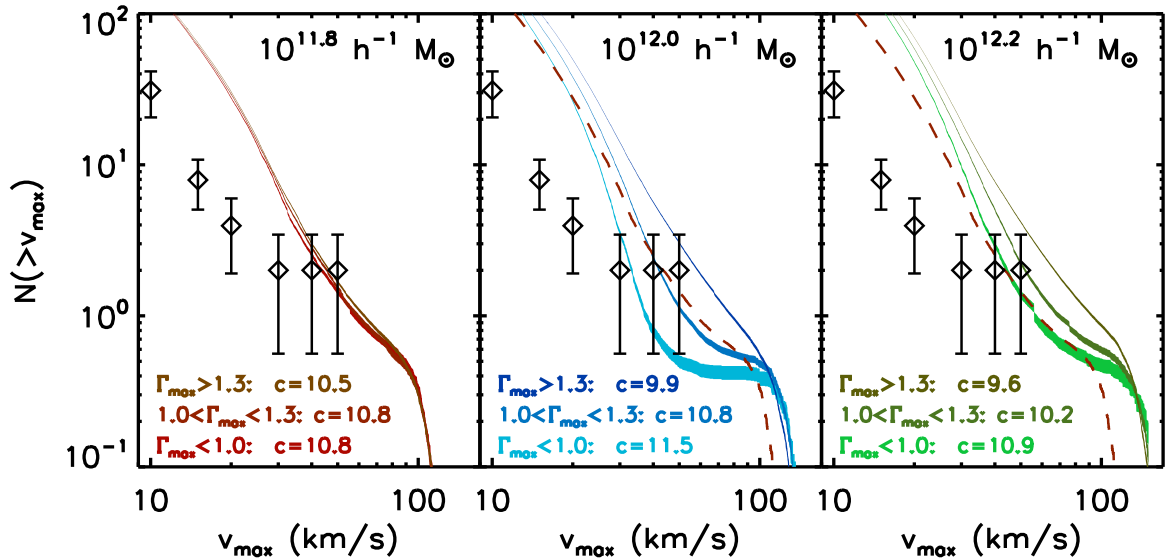}
\caption{The cumulative velocity function $N(>v_{\mathrm{max}})$ describing the number of subhalos with circular velocity 
values larger than $v_{\mathrm{max}}$, as in typical ``missing satellite'' analyses; all {\em solid} curves represent the analytic 
results, where the thickness of the line represents twice the Poisson error.   In each halo-mass model, we divide the catalog into 
samples binned by $\Gamma_{\mathrm{max}}$, finding for the two largest host masses that realizations with lower values of the 
density-proxy parameter have fewer overall satellites, especially at the more massive end of the subhalo spectrum (although 
even in this regime of $\Gamma_{\mathrm{max}}$, there are rare halos hosting very large satellites from recent accretion events).  
Note that decreasing $\Gamma_{\mathrm{max}}$ correlates with an increase in the mean value of the host halo's concentration 
parameter $c$, which in all cases has a Poisson error on the mean of less than 0.1).  In the medium- and high-mass panels, we 
show the mean low-mass behavior ({\em dashed red}) to illustrate the possible degeneracy between halo mass and slope of the 
cumulative velocity function of satellites.  For comparison, we overplot the Milky Way satellite velocity function with Poisson error-bars 
as presented by \cite{simon_geha07}, noting that the low-mass data points likely underestimate the true Galactic dwarf distribution by 
a factor of $\sim 5$ or more, due to luminosity-bias and sky-coverage issues as elaborated by \cite{tollerud_etal08}. }
\label{fig:nvfn}
\end{figure}

Convolving this analysis with the fact that massive minor mergers are relatively rare in Galaxy-scale halos at the 
present-day, although they appear to be occurring in triplicate for the Milky Way today (in the forms of the disrupting 
Sagittarius dwarf and infalling Magellanic Clouds), we have a sense now not only of the discriminative power in 
$\Gamma_{\mathrm{max}}$, but also of the subtlety involved in characterizing halo formation histories by the properties 
of their most massive subhalos.  To avoid contaminating the sample of classical dwarf satellites with these
three ongoing accretion events in the Milky Way, the analyses of \cite{BK_etal11} and related investigations removed 
all subhalos in the VL2 and Aquarius catalogs with $v_{\mathrm{max}} > 80$~km/s at the present day.  
With a large statistical sample of realizations, we can specifically remove the three most massive satellites 
from each realization, thus masking on average the three most recent Galactic accretion events 
by not allowing them to set $\Gamma_{\mathrm{max}}$ in any host halo.  The Magellanic Clouds and Sagittarius are 
not included in the \cite{BK_etal11} analysis; however, it is not entirely obvious which subhalos to remove from a 
theoretically-constructed subhalo catalog to account for this effect.  Removing the three most massive subhalos is the most 
optimistic strategy in terms of driving predictions to agree with data, and we adopt this test in light of the above demonstration 
that the ``too big to fail'' problem is a concordance-cosmology result for at least $10\%$ of Galaxy-sized halos without culling 
any satellites from the catalogs at all.  Certainly a more sophisticated method for describing Sagittarius and the Magellanic Clouds 
would be desirable, as would tighter constraints on individual $\Gamma$-values for the observed classical dwarfs, but considerable 
uncertainty would still remain regarding the properties of their attendant dark matter halos, and the intrinsic scatter in halo/stellar-mass 
relations further complicates the issue \cite{niederste-ostholt_etal10,behroozi_etal10}.

After trimming each subhalo catalog in this way and recalculating the cumulative probability density, we show in 
the right panel of Fig.~\ref{fig:gamma} that the observed Milky Way satellite population becomes much more likely in the 
analytic distribution, occurring in $\sim 40-75\%$ of the realizations and increasing with decreasing host mass as above.  
For comparison, our adopted samples from the VL2 and Aquarius simulations have 
$\Gamma_{\mathrm{max}} \simeq 1.24$~and $1.34$, respectively (note that the Aquarius value represents the most dense 
satellite in a suite of six host halos with virial masses spanning the range between our two most massive models, and also 
that the VL2 host halo is most similar to our heaviest modeled virial mass).  Stacking our catalogs in multiples of six realizations 
for a more direct comparison to Aquarius, we find the mean $\pm 1\sigma$ range of 
$\Gamma_{\mathrm{max}} = (1.23\pm0.21,1.37\pm0.21,1.51\pm0.18)$~for halo masses 
$M_{\mathrm{host}} = (10^{11.8}, 10^{12.0}, 10^{12.2})~h^{-1} $~M$_{\odot}$ respectively.  We also show in Fig.~\ref{fig:gamma} 
these medium- and high-mass ranges of $\Gamma_{\mathrm{max}}$ in stacks of six realizations, since this mass regime 
is more properly compared to the Aquarius host halos.  Based on this analysis, the values of $\Gamma$ in our subhalo populations 
seem to be broadly consistent with the results of the VL2 and Aquarius simulations, suggesting at least that we are not subject to an 
egregious systematic error.  

A relatively low value of $\Gamma_{\mathrm{max}}$ represents a dearth of dense, massive subhalos in a particular host halo, so it is 
instructive to consider the subhalo populations within low-$\Gamma_{\mathrm{max}}$ hosts.  This is interesting for several reasons.  
First, it may lead toward novel predictions that can be used to test our solution, based on large variation in subhalo populations among 
host halos, of the ``too big to fail'' problem.  Second, it may have some bearing on the long-standing ``missing satellite problem'' 
\cite{klypin_etal99,moore_etal99}.  In Figure~\ref{fig:nvfn}, we show the cumulative velocity functions $N(>v_{\mathrm{max}})$ describing 
the distribution of satellites in each of our three models, subdividing each host halo mass into three samples binned by the value of 
$\Gamma_{\mathrm{max}}$ of each realization.  In the two largest host-mass models, the subsets with progressively smaller 
$\Gamma_{\mathrm{max}}$ have steeper $N(>v_{\mathrm{max}})$ functions with lower overall normalizations, which also correspond to 
samples with progressively larger host halo concentration values.  The variation between subsets is much less pronounced in the low-mass 
$M_{\mathrm{host}} = 10^{11.8}~h^{-1} $~M$_{\odot}$ model, in which the host concentration values are similar across subsets of 
$\Gamma_{\mathrm{max}}$.  For comparison, we show the velocity function of Milky Way dwarfs (including those satellites typically called 
``ultra-faint'') as obtained by ref.~\cite{simon_geha07}, noting that the low-mass end of this data set is subject to the sky-coverage and 
luminosity-bias issues elaborated by ref.~\cite{tollerud_etal08} among others.

 \section{Discussion}
\label{sec:discuss}

Using stellar kinematical data, ref.~\cite{BK_etal11} compared constraints on the structural properties of the halos that 
host the dwarf satellite galaxies of the Milky Way with the structural parameters predicted by a number of 
numerical simulations of the formation of Milky Way-sized dark matter halos.  In doing so, they identified a 
discrepancy: too many numerical subhalos have $v_{\mathrm{max}}$ values larger than would be expected 
for systems of their size, when set against these Galactic observations.  If $\Lambda$CDM cosmological predictions 
and/or astrophysical models are not too badly wrong, then do massive invisible dwarfs exist around the Milky Way?  
How could such putative objects have failed to light up with star formation?  The authors of ref.~\cite{BK_etal11} dubbed 
this issue the ``too big to fail" problem.

Soon afterward, ref.~\cite{wang_etal12} showed that one way to mitigate the ``too big to fail'' discrepancy is for the 
Milky Way to reside within a host dark matter halo at the lower-mass range of contemporary constraints 
on the size of the Milky Way halo.  Specifically, this investigation of the Millennium Simulation statistics found that 
for a threshold $v_{\mathrm{max}} = 30$~km/s, fully $\sim 40\%$ of halos with host mass 
$M_{\mathrm{host}} = 10^{12} M_{\odot} =10^{11.85}~h^{-1} $~M$_{\odot}$ host three or fewer satellites larger than the 
threshold (compared to the eight such satellites typical of the Aquarius halos).  For halos more massive than
$M_{\mathrm{host}} \gtrsim 2 \times 10^{12} M_{\odot} =10^{12.15}~h^{-1} $~M$_{\odot}$, this probability drops below $5\%$ 
and quickly vanishes thereafter with increasing halo mass.  Our results are certainly in broad agreement with this conclusion, 
although the two predictions do not probe identical probability distributions; the summary statistic $\Gamma$ involves 
$r_{\mathrm{max}}$ as well as the $v_{\mathrm{max}}$ distribution probed by ref.~\cite{wang_etal12}.    
We therefore focus on the parameter $\Gamma$ because it can directly address the degeneracies in halo properties 
permitted by the data, while noting that a simple translation of our $\Gamma_{\mathrm{max}} < 1$ criterion into 
$v_{\mathrm{max}} < 30$~km/s results in $\sim37\%$ of $M_{\mathrm{host}} = 10^{11.8} $~M$_{\odot}$ hosts satisfying the 
constraints, in significant agreement with the sample analyzed by ref.~\cite{wang_etal12}.

Proposed solutions to the  ``too big to fail" problem, in looking for a means to reduce subhalo internal density, have ranged 
from the cosmological (if satellite centers are scoured by self-interacting dark matter as in ref.~\cite{vogelsberger_etal12}) 
to the astrophysical (if dwarf galaxies sweep out dark mass via feedback effects as in refs.~\cite{diCintio_etal11,BK_etal12,zolotov_etal12}).
In this paper, we explore an additional way in which the ``too big to fail'' discrepancy may be 
mitigated.  In particular, we propose that the large variation in subhalo populations among different 
host halos can explain the dearth of large, dense subhalos orbiting the Milky Way without any making 
any adjustments to the host halo mass or accounting for baryonic feedback processes.  For a host 
halos at the high-mass end of the generally accepted range for the host halo of the Milky Way, 
$M \approx 10^{12.2}\, h^{-1}$~M$_{\odot}$, we find that at least $\sim 10\%$ of all host halos would 
harbor a subhalo population consistent with the observed stellar kinematics of the Milky Way 
dwarfs.  This probability can be considerably higher if the three objects neglected in the 
\cite{BK_etal11} analysis (Sagittarius and the Magellanic Clouds) exhibit any of the highest 
densities among the Milky Way subhalo population.  
  
In formulating our comparison of theoretical predictions with observational constraints, 
we have introduced a new density-proxy parameter $\Gamma$, that roughly runs perpendicular 
to the degeneracy between the subhalo structural parameters $v_{\mathrm{max}}$ and $r_{\mathrm{max}}$ 
exhibited by the data.  Our detailed results are statements about the approximate probability 
distribution of the parameter $\Gamma$ among subhalos in different hosts.  Our study has not identified 
additional diagnostics that can be used to characterize the subset of host halo merger histories that produce 
satellite populations not subject to the ``too big to fail'' problem.  From Fig.~\ref{fig:gamma}, we see that enforcing 
a recent epoch of relative quiescence on the host halo does not significantly increase probability for harboring an 
acceptable subhalo population.  The one exception to this conclusion is for the lowest host halo mass, 
$M_{\mathrm{host}} = 10^{11.8}\,h^{-1}$~M$_{\odot}$.  The probability increase reflects the fact that less massive 
halos are more likely to have experienced minor mergers of a given mass ratio by the present day \cite{stewart_etal08}, 
and this relative homogeneity among host halos also manifests in the cumulative velocity functions and average 
concentration parameters $c$ shown in Fig.~\ref{fig:nvfn}, which do not vary outside the Poisson error on the mean 
for the $M_{\mathrm{host}} = 10^{11.8}\,h^{-1}$~M$_{\odot}$ model.  

Using the large variance of available merger histories to test the observational consistency of any particular 
realization's subhalo population could be an avenue of potential interest in an era of near-field cosmology 
that seeks to connect specific substructure in the Milky Way to characteristics of Galaxy-formation theory.  
Our rough test along those lines in the present work differentiates mainly between systems that {\em have} 
undergone their allotted minor mergers versus those that {\em have not yet} been impacted by high-mass 
subhalos; in a sense, we already know that the Galaxy is part of the latter class, due to the unusually-thin and 
dynamically-cold state of the stellar disk, and that in fact it is presently undergoing multiple minor mergers.  
However, to get a much more detailed handle on the assembly history of the Milky Way by inspecting
analytic distributions of $\Gamma$ and $\Gamma_{\mathrm{max}}$, one would require a reasonable 
estimate of $v_{\mathrm{max}}$ for each individual satellite galaxy's host halo, and it is unclear what (if any) 
fundamental science could be done with such an effort, beyond refining the correspondence between 
numerical results, theoretical predictions, and observational constraints (which we argue here to be 
statistically sound, in contrast to the claims implicit to the ``too big to fail" conjecture).

Fig.~\ref{fig:nvfn} shows subtle distinctions between the properties of host systems at the two 
larger masses ($M_{\mathrm{host}}=10^{12.0}\,h^{-1}$~M$_{\odot}$ and $M_{\mathrm{host}}=10^{12.2}\,h^{-1}$~M$_{\odot}$) 
that pass the ``too big to fail'' test and those that do not.  Hosts halos of these masses that contain subhalo 
populations that are grossly consistent with the Milky Way satellite populations also generally have 
concentrations that are slightly higher than average, though this systematic offset is smaller than 
the dispersion in concentration at fixed mass \cite{bullock_etal01,neto_etal07,maccio_etal07}.  
This weak trend is broadly consistent with the sense that the Milky Way may be an unusually small disk galaxy 
among similarly-sized systems, in terms of stellar mass as well as angular momentum \cite{hammer_etal07}, 
as $\Lambda$CDM models predict that disk size increases with decreasing host halo concentration \cite{bullock_etal01}.  
This implication also appears to be consistent with recent measurements that correlate satellite concentration with 
host stellar mass and total halo mass, in nearby disk-dominated galaxies \cite{cibinel_etal12}.  

A possibly more dramatic feature of subhalo populations within host halos in our higher-mass samples that 
are not in conflict with Milky Way data is that they exhibit relatively steeper cumulative velocity functions, 
as shown in Fig.~\ref{fig:nvfn}.  At best, this only slightly ameliorates the discrepancy that has come to be known as 
the ``missing satellite'' problem, if we take low $\Gamma_{\mathrm{max}}$ samples to 
represent possible Milky Way realizations.  However, as large-scale surveys of the 
Galactic environment complete the local census of dwarf satellites, the faint-end slope may probe 
this prediction of models that solve the ``too big to fail'' problem by exploiting the 
large variance in subhalo populations.  We note, however, that Fig.~\ref{fig:nvfn} appears to indicate that halo mass 
and velocity-function slope may be somewhat degenerate, and therefore a full exploration of this parameter space 
by observations of substructure in extragalactic systems could be useful in discriminating between the competing effects 
we discern here.

Interestingly, the steeply-sloped realizations with low $\Gamma_{\mathrm{max}}$ also display a plateau beneath a value of 
unity in the subhalo velocity function.  This reflects the fact that some relatively rare Galaxy-sized halos may have a very massive 
companion with $v_{\mathrm{max}} > 60$~km/s, with such a subhalo having necessarily fallen in recently (since dynamical friction destroys 
massive objects quickly).  We note that the Milky Way dwarf population mimics this behavior somewhat, within observational Poisson 
error of the analytic plateau, due to the presence of three large satellites with $v_{\mathrm{max}} > 60$~km/s as well as the absence of any 
moderately-massive subhalos between $30 < v_{\mathrm{max}} < 60$~km/s, supporting the conclusion that the low $\Gamma_{\mathrm{max}}$ 
description of the Galactic dwarfs is consistent with a flat velocity function in this intermediate range, and favors the presence of one or possibly 
more large systems with relatively higher $v_{\mathrm{max}}$.  In the particular case of the Milky Way, this plateau occurs at three large satellites, 
due to the conjoined Magellanic Clouds and the currently-disrupting Sagittarius dwarf galaxy.

In summation, we have used analytic models of subhalo populations to argue that the 
absence of luminous Milky Way satellite galaxies residing in halos of high density 
may not indicate a problem with contemporary theories of structure formation or 
galaxy formation.  Rather, we have argued that the non-negligible variation in 
subhalo populations may reasonably account for this deficit.  We estimate that 
at least 10\% of halos could be consistent with observations without any requirement 
that the host halo of the Milky Way have a mass toward the lower range of contemporary 
constraints.  There are several implications of this result.  First, it suggests that there is no 
need to consider relatively low-mass hosts for the Milky Way (though such may certainly be 
the case, and could be a large part of the solution).  Second, it shows that the ``too big to fail'' 
problem cannot be a statement of high statistical significance.  Further study is necessary 
to confirm or refute our argument.  In particular, we have, by necessity, used 
approximate techniques to predict the properties of dark matter subhalos so a 
large numerical simulation campaign will be necessary to test our predictions.  
In performing such a follow-up study, it must be borne in mind that selecting 
particular halos for high-resolution resimulation using accretion history 
or halo structural information \cite{springel_etal08,diemand_etal07}
can introduce biases; however, we have not been able to identify clear biases 
in our analytic models.  Furthermore, we have neglected the effects of baryons 
and these may need to be considered more carefully.  Finally, once a large 
set of simulated subhalo populations is available, it will be imperative to 
conduct a full statistical comparison of theoretical predictions with 
stellar kinematical data.  Such follow-up studies may, indeed refute our 
argument perhaps by identifying undiagnosed systematic errors in our methods.  
However, absent such detailed follow-up, the ``too big to fail'' problem 
is unlikely to pose a serious challenge to the standard, hierarchical cold dark 
matter model of structure growth; rather than being a potential wound to the predictive 
power of $\Lambda$CDM, it may instead represent an exhibition of its productive variety.  

\acknowledgments{We would like to thank Michael Boylan-Kolchin, Andrew Hearin, and Mike Kuhlen
for useful discussions.  This work was funded by the University of Pittsburgh, by the 
Pittsburgh Particle physics, Astrophysics, and Cosmology Center 
(PITT PACC), and by the National Science Foundation through Grants NSF PHY 0968888 
and NSF AST 1108802.}

\bibliographystyle{jhep}
\bibliography{bailing}

\end{document}